\documentstyle[multicol,aps,prl,graphics]{revtex}
\draft
\begin{document}

\newcommand{\be}{\begin{equation}}
\newcommand{\ee}{\end{equation}}
\newcommand{\bea}{\begin{eqnarray}}
\newcommand{\eea}{\end{eqnarray}}
\newcommand{\beb}{\begin{eqnarray*}}
\newcommand{\eeb}{\end{eqnarray*}}
\newcommand{\nn}{\nonumber}


\preprint{SNUTP 01-021}
\title{Synchronization on small-world networks}
\author{H. Hong and M.Y. Choi}
\address{
Department of Physics and Center for Theoretical Physics, 
Seoul National University, 
Seoul 151-747, Korea}
\author{Beom Jun Kim}
\address{
Department of Theoretical Physics, Ume{\aa} University, 901 87 Ume{\aa}, Sweden 
}

\maketitle

\begin{abstract}
We investigate collective synchronization in a system of coupled
oscillators on small-world networks.  The order parameters which measure 
synchronization of phases and frequencies are introduced and analyzed 
by means of dynamic simulations and finite-size scaling.  
Phase synchronization is observed to emerge in the presence of even a tiny 
fraction $P$ of shortcuts and to display saturated behavior for $P \gtrsim 0.5$. 
This indicates that the same synchronizability as the random network ($P=1$)
can be achieved with relatively small number of shortcuts. 
The transient behavior of the synchronization, obtained from the measurement of the
relaxation time, is also discussed. 
\end{abstract}


\begin{multicols}{2}
\pagebreak

Systems of coupled nonlinear oscillators, which serve as prototype models 
for various oscillatory systems in nature, have attracted much attention.
Those systems exhibit remarkable phenomena of collective synchronization,
which have been observed in a variety of physical,
biological, and chemical systems~\cite{ref:synch}.
Up to date, existing studies on collective synchronization have mostly 
been performed either on the local regular networks such as $d$-dimensional cubic 
lattices or on the globally connected geometry.
In recent years, there has been suggested the possibility that
a number of diverse systems in nature may have the same topological structure
as the small-world networks~\cite{ref:WS},
which are intermediate of the local regular networks and the fully random networks.
Such small-world networks are usually characterized by
two interesting features: high clustering, which is a characteristic
of regular networks, and the short path length, which is typically
observed in random networks~\cite{ref:WS}.
Most studies on small-world networks have been focused on the
geometrical and topological characterization of the networks,
with little attention paid to dynamics defined on them. 
Recently, some studies have considered dynamical systems put on
small-world networks~\cite{ref:Watts,ref:NHmodel}, where such desirable features
as faster propagation of information, better computational power,
and stronger synchronizability have been observed.
In Ref.~\cite{ref:Watts}, frequency synchronization on the small-world 
network has been noticed in the presence of a small amount of randomly rewired 
connections 
and the possibility of the transition to global entrainment 
with the mean-field nature has been pointed out. 
However, quantitative analysis has not been performed and proper understanding
is still lacking.  For example,
the critical rewiring probability beyond which true long-range order is present
at finite coupling strength has not been addressed. 

In this paper we study the detailed aspects
of the collective synchronizations on small-world networks,
as the rewiring probability and the coupling strength are varied.
In general, frequency synchronization can be attained without 
synchronization of phases, and we explore both 
to investigate the synchronization-desynchronization transition. 
Via careful finite-size scaling, we find the following:
(i) Phase synchronization as well as frequency one, which is absent in one-dimensional 
regular networks, emerges in the presence of even a very small fraction of shortcuts. 
(ii) The phase synchronization transition is of the mean-field type, the same as the
Kuramoto model~\cite{ref:Kuramoto}.  (iii) The relaxation time monotonically
decreases with the rewiring probability $P$ up to $P=0.5$ 
and apparently saturates for $P\gtrsim 0.5$.
This indicates that the time 
required to synchronize the small-world network for $P \gtrsim 0.5$
is almost the same as that for a random network ($P=1$).

According to Ref.~\onlinecite{ref:WS}, the small-world network is 
constructed in the following way: 
First, a one-dimensional regular network with only local connections 
(of range $k$) between the $N$ nodes is constructed.  
Each local link is visited once, and then with the rewiring probability $P$ 
it is removed and reconnected to a randomly chosen node. 
After the whole sweep of the entire network, the total number of shortcuts
in the network is given by $NPk$ for sufficiently large $N$. 
At each node of this small-world network is located an oscillator;
a link connecting two nodes represents coupling between the 
two oscillators at those two nodes.
Describing the state of the $i$th oscillator, i.e., the one at node $i$ by
its phase $\phi_i$, we write the set of equations of motion governing the dynamics 
of the $N$ oscillator system $(i= 1,2,\cdots,N)$~\cite{ref:Kuramoto}: 
\be \label{eq:model}
\dot{\phi_i }(t)
+ \frac{K}{2k} \sum_{j \in \Lambda_i} \sin (\phi_i -\phi_j )
= \omega_i ,
\ee
where $\Lambda_i$ denotes the set of nodes connected to node $i$ 
(via either local links or shortcuts)
and $K$ is the coupling strength suitably normalized with respect to 
the average number of connections per node. 
On the right-hand side $\omega_i$ represents the intrinsic frequency 
of the $i$th oscillator.  They are quenched random variables with the distribution 
function $g(\omega)$.

On the small-world network built in this manner with given $k$ and $P$,
we investigate the collective synchronization behavior 
of the coupled oscillators at various values of the coupling strength $K$.
For convenience, the range $k=3$ is taken and 
the Gaussian distribution with unit variance ($\sigma^2 =1$) for $g(\omega)$ 
is used~\cite{foot:g}. 
We then use the Heun's method~\cite{Heun} 
with the discrete time step $\Delta t =0.05$, to integrate numerically
Eq.~(\ref{eq:model}). 
Typically, while the equations of motion are integrated for
$N_t = 4\times 10^3$ time steps, the data from the first $N_t/2$ steps
are discarded in measuring quantities of interest. 
Both $\Delta t$ and $N_t$ have been varied to verify that the measured
quantities are precise enough and the networks of various sizes, up to $N=3200$,
have been considered. 
For each network size, we have performed one hundred independent runs with
different configurations of the intrinsic frequencies 
as well as different network realizations, over which averages have been taken. 

Collective behavior of the oscillator system is conveniently described by the
order parameters
\begin{eqnarray}
m &\equiv& \left[\left\langle \Biggl| \frac{1}{N}\sum_{j=1}^{N} e^{i\phi_j} \Biggr| \right\rangle \right],  \label{eq:m} \\
q &\equiv& \left[\left\langle \frac{2}{N(N-1)}\sum_{i<j}^{N}
e^{-c(\dot\phi_i -\dot\phi_j )^2}\right\rangle \right], \label{eq:q}
\end{eqnarray}
where $\langle \cdots \rangle$ and $[ \cdots ]$ denote the averages
over time and over different realizations of the intrinsic frequencies, 
respectively, and $c$ is a sufficiently large number. 
Since the frequency resolution is given by $(N_t \Delta t)^{-1}$ 
in numerical simulations, 
two oscillators should be regarded as mutually entrained if the difference in 
frequency is smaller than $(N_t \Delta t)^{-1}$. 
The order parameter $q$ in Eq. (\ref{eq:q}) does not depend sensitively on 
the value of $c$, 
and we thus choose $c =10^6 $, which turns out to be sufficient
for perceiving the difference. 
For comparison, we have also considered a different version of 
the frequency order parameter
$r\equiv$ \(\lim_{N\rightarrow\infty} (N_s/N)\), where $N_s$ is the 
number of mutually entrained oscillators in the largest cluster~\cite{ref:freqorder}, 
only to find no essential difference.

\begin{figure}
\resizebox*{!}{5.7cm}{\includegraphics{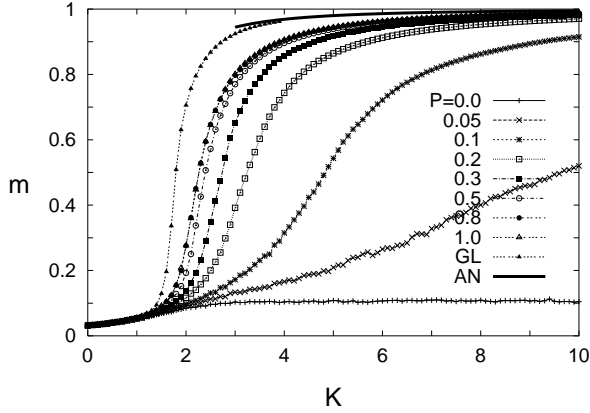}}
\caption{Phase synchronization order parameter $m$ as a function of the
coupling strength $K$ for various values of the rewiring probability $P$
in the small-world network with size $N=800$.  
For comparison, obtained data and known analytic results 
for the globally-connected network, labeled as ``GL'' and ``AN'' respectively, are 
also plotted.  The error bars estimated by the standard deviation have  
approximately the sizes of the symbols and the lines are merely guides to the eye. 
It is shown that phase synchronization appears at all nonzero values of $P$.
}
\label{fig:phasesynch}
\end{figure}

\begin{figure}
\resizebox*{!}{5.7cm}{\includegraphics{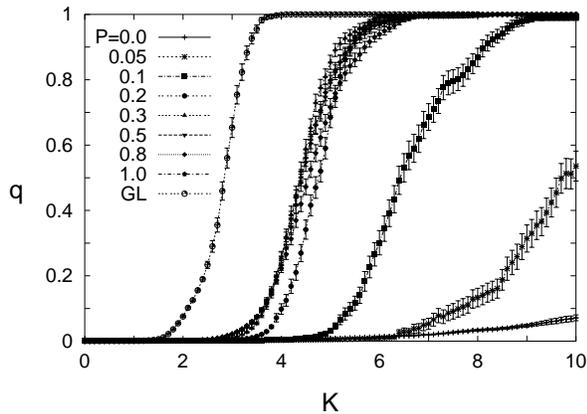}}
\caption{Frequency synchronization order parameter $q$ versus $K$ for
various values of $P$ in the small-world network with size $N=800$.  
The results of the globally-connected network are also plotted for comparison.  
Error bars represent standard deviations and lines are guides to the eye.
Frequency synchronization is shown to emerge at all nonzero values of $P$.
}
\label{fig:freqsynch}
\end{figure}

Figures~\ref{fig:phasesynch} and \ref{fig:freqsynch} display
the obtained phase and frequency order parameters ($m$ and $q$)
versus the coupling strength $K$ at various values of the rewiring probability.
In the weak coupling limit ($K \rightarrow 0$), phases of the oscillators are 
distributed uniformly on the interval $[0, 2\pi ]$, 
yielding $m = {\cal O}(1/\sqrt{N})$ and the absence of macroscopic coherence.
On the other hand, in the strong coupling limit ($K \rightarrow \infty$)
all phases of the oscillators become synchronized, to give $m=1$ and 
accordingly $q=1$, regardless of the detailed structure of the 
network~\cite{foot:mq}.

An important observation from Figs.~\ref{fig:phasesynch} and
\ref{fig:freqsynch} is that synchronization of the phase 
as well as of the frequency 
exhibits strong dependence on the rewiring probability $P$.  
In particular, both of them do not show synchronization
in the absence of shortcuts ($P=0$), which is consistent with the known result 
in a one-dimensional system~\cite{ref:freqorder}.  
When a tiny fraction of
the shortcuts comes into the system, on the other hand, 
the dynamics of the system changes dramatically, 
giving rise to phase and frequency synchronization 
(compare curves for $P=0$ and for $P=0.05$ 
in Fig.~\ref{fig:phasesynch} and \ref{fig:freqsynch}). 
Another interesting feature found in Fig.~\ref{fig:phasesynch} is that 
as $P$ grows, phase synchronization, measured by $m(K)$ in Fig.~\ref{fig:phasesynch}, 
saturates and does not show significant difference for $P > P_m \approx 0.5$. 
It is thus indicated that phase synchronization almost the same as 
that for $P=1.0$ can be achieved with relatively small amount of 
shortcuts ($P \approx 0.5$).  
This may have some practical importance in network systems where
making long-ranged shortcuts has a high cost of resources.  
>From an economical point of view, small-world networks with $P \approx P_m $ 
is favorable: Global coherence is attained with resources spent as less as possible.
Similar behavior may also be expected for the frequency synchronization measured 
by $q(K)$ in Fig.~\ref{fig:freqsynch}, 
although rather large finite-size effects, particularly for small $P$, 
tend to obscure such saturation behavior of $q$. 

For comparison, the synchronization behavior in the globally-connected network 
is also displayed in Figs.~\ref{fig:phasesynch} and \ref{fig:freqsynch},
where good agreement with the analytic expression $m \approx 1 - \sigma^2/2K^2$
found in Ref.~\onlinecite{ref:spontaneous} for large $K$ 
is shown (see the thick solid line labeled by ``AN'' in Fig.~\ref{fig:phasesynch}). 
Noting that the number of connections in a small-world network is ${\cal O}(N)$,
in sharp contrast with ${\cal O}(N^2)$ valid in the globally-connected network, 
we naturally expect that the phase synchronization on a small-world
network sets in at the coupling strength larger than the critical value 
$K_c = 2/\pi g(0) \approx 1.60$ in a globally-connected network
[for the Gaussian distribution $g(\omega)$ with unit variance]~\cite{ref:Kuramoto}. 
Indeed such features are shown in 
Figs.~\ref{fig:phasesynch} and \ref{fig:freqsynch}. 
Furthermore, it is noteworthy that qualitatively the same 
synchronization behavior is observed for
the globally-coupled network and the small-world network, 
which implies that strong synchronization can be achieved with only ${\cal O}(N)$
connections instead of ${\cal O}(N^2)$. 

Precise determination of the critical coupling strength $K_c$ separating
desynchronized and synchronized states requires careful consideration
of the finite-size effects.  
We examine finite-size scaling of the 
phase order parameter to determine $K_c$ and explore the transition nature 
around it.
In the thermodynamic limit the order parameter displays the critical behavior 
\begin{equation} \label{eq:mbeta}
m \sim (K - K_c )^{\beta},
\end{equation}
with the critical exponent $\beta$.
On the other hand, in a finite system, we expect 
$m = (K - K_c )^{\beta} f(\xi/N, \zeta/N)$ with a function $f$
of two scaling arguments $\xi/N$ and $\zeta/N$, where the 
correlation length $\xi$ diverges at $K_c$, 
and $\zeta \,(\equiv 1/kP)$ is an 
additional length scale corresponding to the typical distance between the 
ends of shortcuts in the small-world network~\cite{ref:zeta}. 
Here we pay attention to the system with size much larger than 
$\zeta$, and approximate the above scaling function $f(\xi/N, \zeta/N)$ 
as $f(\xi/N, 0)$. 
This leads to the scaling form of the 
order parameter 
\begin{equation} \label{eq:F}
m = N^{-\beta/{\bar\nu}} F\left((K - K_c)N^{1/{\bar\nu}}\right),
\end{equation}
where the critical exponent $\bar\nu$
describes the divergence of the 
correlation volume $\xi_V$ at $K_c$~\cite{foot:xi,ref:xi}:
\begin{equation} \label{eq:correlation}
\xi_V \sim |K - K_c |^{-\bar\nu }. 
\end{equation}
Since at $K=K_c$ the function $F$ in Eq.~(\ref{eq:F}) has a value
independent of $N$, one can determine $K_c$ by means of the standard 
finite-size scaling analysis. 
Namely, plotting $mN^{\beta/{\bar \nu}}$ versus $K$ for various
sizes, one can find the value of $\beta/{\bar \nu}$ which gives
a well-defined crossing point at $K_c$.  
After $\beta/{\bar \nu}$  and $K_c$ are determined, one then use 
\begin{equation}
\ln\left[\frac{d m}{dK}\right]_{K_c} =  \frac{1-\beta}
{\bar\nu}\,\ln N + \mbox{const},
\label{eq:slope}
\end{equation}
in order to obtain the value of $(1-\beta)/{\bar \nu}$, which, combined with
the known value of $\beta/{\bar \nu}$, gives the values of
$\beta$ and ${\bar \nu}$.

\begin{figure}
\resizebox*{!}{5.7cm}{\includegraphics{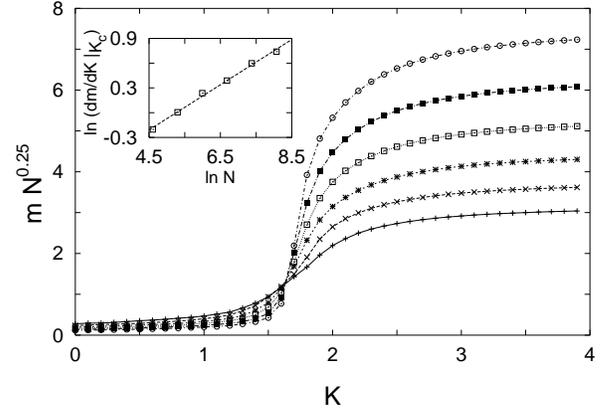}}
\caption{
Phase synchronization order parameter $m$ for the globally
coupled network plotted as 
$m N^{\beta/\bar\nu}$ with $\beta/\bar\nu = 0.25$ versus $K$
for the network size $N=100, 200, 400, 800, 1600$, and 3200 (from bottom to top 
on the right side).  
There is given a unique crossing point at $K_c \approx 1.61$.
Inset: From Eq.~(\ref{eq:slope}), the slope $(1-\beta)/{\bar \nu} \approx 0.27$
is obtained, which, combined with $\beta/\bar\nu \approx 0.25$ 
found in the main panel, results in $\beta \approx 0.48$ and 
$\bar\nu \approx 1.92$. 
} 
\label{fig:GL}
\end{figure}

\begin{figure}
\resizebox*{!}{5.7cm}{\includegraphics{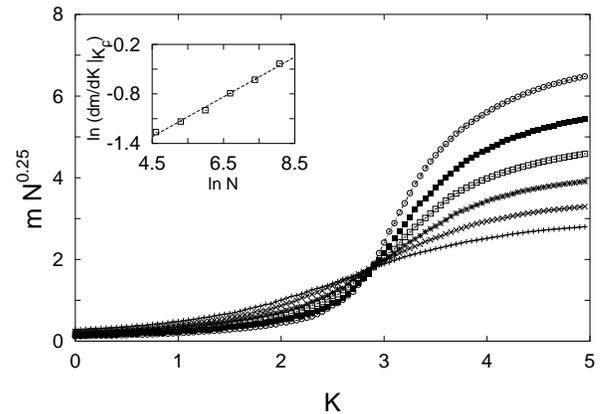}}
\caption{Phase synchronization order parameter 
$m$ for the small-world network with $P=0.2$ plotted as 
$m N^{\beta/\bar\nu}$ with $\beta/\bar\nu = 0.25$ versus $K$, displaying 
one unique crossing point at $K_c \approx 2.88$ 
[$N=100, 200, 400, 800, 1600$, and 3200 from bottom to top 
on the right side].
Inset: $(1-\beta)/{\bar \nu} \approx 0.24$ is found
from the slope.
Analysis similar to that in Fig.~\ref{fig:GL} leads to the same mean-field
exponents $\beta \approx 0.5$ and $\bar\nu \approx 2.0$.
} 
\label{fig:P0.2}
\end{figure}

Figure~\ref{fig:GL} shows the determination of $K_c$ for 
the globally-connected network through the use of the finite-size scaling
form in Eq.~(\ref{eq:F}). 
Varying the value of $\beta/\bar\nu$, we find that $\beta/\bar\nu \approx 0.25$
gives the well-defined crossing point at $K_c \approx 1.61$, which
is in a good agreement with the analytical result 
$K_c = 2/\pi g(0) \approx 1.60$~\cite{ref:Kuramoto}.
In the inset of Fig.~\ref{fig:GL}, the least-square fit to Eq.~(\ref{eq:slope})
gives $(1-\beta)/\bar\nu \approx 0.27$,
which, combined with $\beta/\bar\nu \approx 0.25$, 
yields $\beta \approx 0.48$ and $\bar \nu \approx 1.92$. 
These results are certainly consistent with the fact that 
the globally-connected network is a mean-field system,
which has $\beta = 1/2$ and $\nu =1/2$~\cite{ref:Kuramoto}; in particular
the obtained value of $\bar\nu$ close to $2$ indicates
that the upper critical dimension of the synchronization transition
is four~\cite{foot:xi}.  
Similarly, the transition behavior on the small-world network may be 
investigated.  Shown in Fig.~\ref{fig:P0.2} is the determination
of exponents and $K_c$ for the small-world network with the
rewiring probability $P = 0.2$. 
From the same analysis as in Fig.~\ref{fig:GL}, we obtain 
$K_c \approx 2.88$ together with exponents
$\beta \approx 0.51$ and $\bar\nu \approx 2.04$, 
which are essentially the same as those in the globally-coupled network. 
This concludes that the coupled oscillator system on a small-world network 
with the number of connections given by ${\cal O}(N)$ 
displays a mean-field synchronization transition,
like the system on a globally connected network 
with the much larger number of connections ${\cal O}(N^2)$.

\begin{figure}
\resizebox*{!}{5.7cm}{\includegraphics{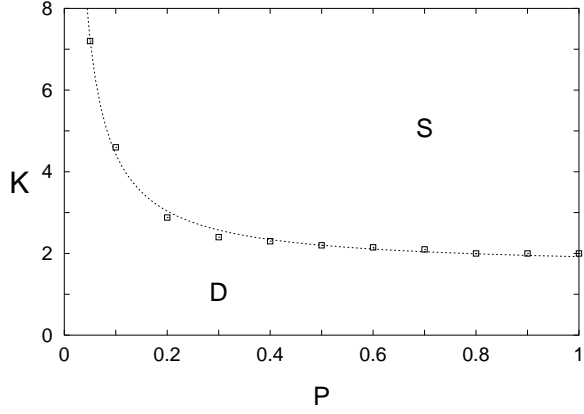}}
\caption{Phase diagram of the oscillator system on a small-world network.  
The data points on the phase boundary have been obtained from the finite-size 
scaling form of the order parameter in Eq.~(\ref{eq:F}).  The phase boundary, 
separating the synchronized (S) state from the desynchronized (D) one, is
well described by the equation $K_c = 1.64(4)+0.28(1) P^{-1}$, 
represented by the dotted line.
}
\label{fig:phd}
\end{figure}

Figure~\ref{fig:phd} displays the phase diagram on the plane
of $K$ and $P$.  The data points on the phase boundary
separating synchronized (S) and desynchronized (D) states
have been obtained from the finite-size scaling analysis described above. 
The nature of the synchronization transition has always been
found to be the mean-field type for all values of $P$ used
in this work.
We find that the phase boundary in Fig.~\ref{fig:phd}
is well described by the equation $K_c = 1.64(4) + 0.28(1)P^{-1}$. 
As $P\rightarrow \infty$, this form predicts $K_c = 1.64(4)$, in good agreement 
with $K_c \approx 1.61$ found in Fig.~\ref{fig:GL} and the known
analytic result $K_c = 2/\pi g(0) \approx 1.60$ for the 
globally-coupled network.  
This agreement is somehow expected since the globally-coupled network,
where the number of long-ranged connections (shortcuts) is ${\cal O}(N^2)$,
corresponds to $P \approx N$. 
Straightforward extrapolation of the estimated form of the
phase boundary predicts that $K_c$ is finite only if $P\neq 0$. 
Assuming its validity, one expects to have synchronization on the small-world
network at finite coupling strength unless $P$ is zero. 
This has close resemblance to the existing
studies on the small-world phenomena: The characteristic
path-length in the small-world network behaves very differently 
for $P=0$ and for $P \neq 0$~\cite{ref:Newman}. 
Thus supported is the view that the small-world transition and 
the order-disorder transition are intimately related~\cite{ref:XY}. 

Meanwhile, the frequency order parameter $q$ is observed to
shift toward larger values of the coupling strength $K$ with the increase of
the system size $N$,
sustaining its shape and eventually converging. 
Such peculiar behavior, which appears more conspicuous at small values of $P$,
makes the standard finite-size scaling analysis inadequate, 
leaving the precise value of the critical coupling strength 
for the frequency synchronization rather difficult to determine. 
In particular it is not clear whether phase synchronization and frequency 
synchronization emerge simultaneously (at the same critical coupling strength)
for all values of the rewiring probability $P$. 
The rather limited data appear to favor 
a slightly smaller value of the critical coupling strength 
for frequency synchronization (at small values of $P$), 
suggesting the interesting possibility of frequency synchronization
without phase synchronization for intermediate values of the coupling $K$. 
For conclusive results, however, more extensive simulations are required. 

\begin{figure}
\resizebox*{!}{5.7cm}{\includegraphics{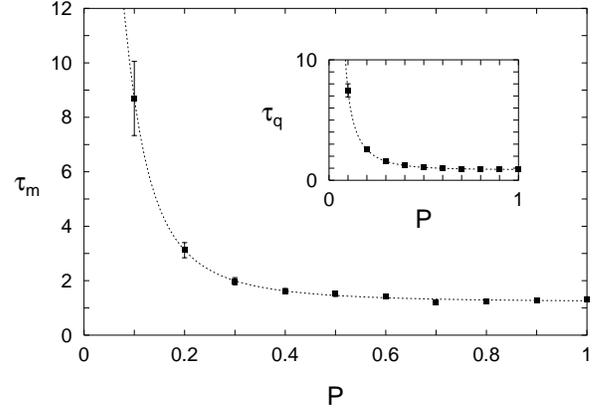}}
\caption{Relaxation time $\tau_m$ of the phase synchronization order
parameter (in arbitrary units) versus $P$ for $K=10$.  
It is manifested that $\tau_m \rightarrow \infty$ as $P \rightarrow 0$.
Error bars represent standard deviations and the line is merely a guide to the eye.
Inset: Relaxation time $\tau_q$ for frequency synchronization (in arbitrary units) 
versus $P$ for $K=10$.} 
\label{fig:relaxtime}
\end{figure}

>From the point of view of information transfer, the remarkably
short path length observed in a small-world network may imply
that information flow through all elements in the 
network is quite fast.  In the present work, this can be
rephrased as follows: The time it takes to establish global
synchronization should decrease substantially as small amounts of shortcuts
are introduced.  
To investigate such transient behavior,
we start from random initial conditions and measure
the phase order parameter $m$ as a function of time. 
The average relaxation time~\cite{ref:Anor}
\begin{equation}
\tau_{m} \equiv \int_{0}^{\infty} dt' {\bar m}(t')
\end{equation}
is then computed, where the normalized order parameter 
\begin{equation}
\bar m \equiv \frac{m(t) - m^{eq}}{m_0 - m^{eq}}
\end{equation}
with the initial value $m_0 \equiv m(t{=}0)$ and the equilibrium value 
$m^{eq} \equiv m(t {\rightarrow} \infty)$
satisfies $\bar m (0) =1$ and $\bar m (t{\rightarrow} \infty) = 0$. 
The average relaxation time defined in this way is
very useful for the systems with many relaxation time scales. 
Here, for $P > 0$, we observe that 
$\bar m(t)$ at long times
does not fit well to the exponential decay form based on 
a single relaxation time scale.
In Fig.~\ref{fig:relaxtime} we show the relaxation time $\tau_m$
versus $P$ at $K=10$.  
As $P$ is increased, $\tau_m$ is shown to decrease up to 
$P \approx 0.5$, and then it appears to saturate for $P\gtrsim 0.5$.
The similarity between the phase boundary in Fig.~\ref{fig:phd} and the relaxation 
time in Fig.~\ref{fig:relaxtime} is striking and suggests 
almost the same synchronization behavior 
(in both transient and stationary aspects)
for all values of $P \gtrsim 0.5$.
We have also investigated the relaxation time $\tau_q$ for 
frequency synchronization and found that similar
saturation appears, as shown in the inset of Fig.~\ref{fig:relaxtime}. 
This reflects the overall similarity in achieving both types of synchronization. 

In summary, we have examined collective synchronization in the system of
coupled oscillators on small-world networks. 
Both the phase and frequency synchronization have been found to exhibit strong 
dependence on the rewiring probability $P$, as $P$ is raised from zero.  
In particular, for any nonzero value of $P$ considered here, 
both types of synchronization emerge at finite coupling strengths 
while they are absent at $P=0$.  This apparently suggests that 
the critical value of $P$, below which synchronization does not set in 
at any finite coupling strength, vanishes, although 
more extensive simulations near $P=0$ are necessary for conclusive results. 
The phase boundary for the (phase) synchronization transition, 
which is of the mean-field type at any $P$, has been obtained from 
the finite-size scaling analysis of the phase order parameter. 
The transient behavior of synchronization has also been investigated 
through the relaxation time of the system, i.e., the time taken by the
system in the desynchronized initial state to reach the final synchronized state.
Both the phase diagram and the relaxation time, measuring
the stationary and transient features of synchronization, respectively,
display similar saturation behavior for $P \gtrsim 0.5$. 
>From a practical point of view, this saturation behavior has a useful
implication on building networks: 
Since the long-range connection usually costs more than the local connection,
it is advantageous to tune $P$ to the value at which the saturation begins
and to establish globally connected behavior with less consumption of
resources.
Finally, the possibility of successive transitions with the coupling strength,
one for frequency synchronization and the other for phase synchronization, 
has been raised at small values of $P$, the detailed investigation of which 
is left for future study.  

This work was supported in part by the Ministry of Education of Korea through the BK21 
Program (H.H. and M.Y.C.), and in part by the Swedish Natural Research Council through 
Contract No. F 5102-659/2001 (B.J.K.).

\end{multicols}


\begin{references}
\bibitem{foot:g} We have also considered other types of distribution including
the uniform one, only to find no qualitative difference.
\bibitem{foot:mq} In general phase synchronization ($m=1$) implies frequency
synchronization ($ q = 1$), while the reverse is not true.
\bibitem{foot:xi}
In $d$ dimensions, we have $\xi_{V} \sim \xi^d $ with the correlation length
diverging as $\xi \sim |K - K_c |^{-\nu}$. 
This leads to the relation $\bar\nu = \nu d$ in a $d$-dimensional system; 
in a mean-field system ($\nu =1/2$), we have $\bar\nu = d_c/2$ with the
upper critical dimension $d_c$.
\end{references}
\end{document}